\documentclass[prd,twocolumn,showpcs,amsmath,amssymb,nofootinbib,preprintnumbers,balancelastpage]{revtex4}
\pdfoutput=1
\usepackage{epsfig}
\usepackage{amsmath}
\usepackage{bm}
\usepackage{times}
\usepackage{graphicx}
\usepackage{color}
\usepackage{slashed}
\usepackage{graphicx}
\usepackage{amsmath, amssymb}

\def\bea{\begin{eqnarray}}
\def\eea{\end{eqnarray}}
\def\ba{\begin{eqnarray}}
\def\ea{\end{eqnarray}}
\def\be{\begin{equation}}
\def\ee{\end{equation}}

\begin{document}
\preprint{Published in JHEP}

\title{Standard model with compactified spatial dimensions}

\author{Bartosz Fornal and Mark B. Wise\\
\textit{California Institute of Technology, Pasadena, CA 91125, USA}\\
}
\date{\today}

\begin{abstract}
We analyze the structure of the standard model coupled to gravity
with spatial dimensions compactified on a three-torus.
We find that there are no stable one-dimensional vacua at zero temperature,
although there does exist an unstable vacuum for a particular set of Dirac neutrino masses.
\vspace{11mm}
\end{abstract}

\maketitle
\bigskip

\section{Introduction}
The standard model coupled to gravity has a unique four-dimensional vacuum.
Nevertheless, in case of one spatial
dimension compactified on a circle
\cite{Arkani}, or for two spatial dimensions compactified on a 2-torus, it has  recently been shown that there may also exist lower-dimensional
vacua stabilized by the Casimir energies of the standard model particles with the lowest mass, i.e., gravitons, photons and neutrinos.   Such  vacua of the low-energy effective theory exist at zero temperature for a wide range of
experimentally allowed neutrino masses.
In \cite{AFI}
it was shown that  at high enough temperatures  these stationary points  are washed out.  At zero temperature  an extremely small  rate for tunneling to a lower-dimensional anti-de Sitter spacetime was found following the steps outlined in \cite{Sean}.

This work completes the series of papers \cite{Arkani,AFW,AFI} concerning lower-dimensional standard model vacua
by considering the last remaining case, when all spatial dimensions are compactified.
We analyze the compactifications on $T^3$, $S^1\times S^1\times S^1$, $S^1\times T^2$, $S^3$, and $S^1\times S^2$,
but our primary focus is on the 3-torus case, since it seems the most natural three-dimensional
topology with no curvature.

Three-dimensional compactifications  are qualitatively different from the one- and two-dimensional compactifications, since a stable vacuum cannot occur for a ``generic range'' of neutrino masses. Nonetheless, a brief study of this case is worthwhile.
The geometry of the lower-dimensional vacuum is determined by the shape of the effective potential, which is a sum
of Casimir energies of the particles and the cosmological constant term.
We show that this potential for the 3-torus case has  no stable stationary points  at zero temperature.
For the standard model with Dirac neutrinos, however, there does exist an unstable stationary point
for a particular set of neutrino masses, depending on the type of hierarchy.

\section{Compactification on a 3-torus at zero temperature}
In this section we explore the existence of lower-dimensional vacua of the standard model coupled to gravity with spatial dimensions
compactified on a 3-torus. We start with the 4D Einstein-Hilbert action,
\begin{eqnarray}\label{4Daction}
S = \int d^4x \sqrt{-g}\left[\frac{1}{2}M_{p}^2
{\cal R}+\mathcal{L}_{\rm SM}\right],
\end{eqnarray}
where $g$ is the determinant of the 4D metric, the Planck mass $M_{p}\simeq 2.4 \times 10^{18}~{\rm GeV}$,
$\mathcal{R}$ is the Ricci scalar, and $\mathcal{L}_{\rm SM}$ is the standard model Lagrangian including
the cosmological constant.
Consider the following spacetime interval,
\vspace{2mm}
\bea
ds^2 = - N^2 d t^2 + T_{i j} d y^i d y^j\,,
\eea
where $T_{i j}$ is the metric on the 3-torus with $i, j = 1, 2, 3$ and the compact coordinates $y^i \in [0, 2\pi)$.
We adopt the same parametrization as in \cite{Arkani},
\vspace{2mm}
\bea\label{torus}
T_{i j} = \frac{b^2}{(\rho_3 \tau_2)^{2/3}}\left(
                              \begin{array}{ccc}
                                1 & \tau_1 & \rho_1 \\
                                \tau_1 & \tau_1^2+\tau_2^2 & \rho_1\tau_1 + \rho_2\tau_2 \\
                               \rho_1 & \rho_1\tau_1+ \rho_2\tau_2 & \rho_1^2+\rho_2^2+\rho_3^2 \\
                              \end{array}
                            \right) ,
\eea
where $\Psi^{\rm T} = (\tau_1, \tau_2, \rho_1, \rho_2, \rho_3)$ are the shape moduli and $b^3$ is the volume modulus, all functions only of time.
The dimensionally reduced action is,
\vspace{3mm}
\begin{eqnarray}\label{1Daction}
S = \!\!\int \!d t \!\left[\frac{1}{2}M_{p}^2\frac{(2\pi b)^3}{N}\left(\!-6\,\frac{\dot{b}^2}{b^2} \!+\! \dot{\Psi}^{\rm T} \hat{M} \dot{\Psi}\!
\right)\!-\!N \,V(b, \Psi)\!\right]
\end{eqnarray}
where the dot indicates a derivative with respect to time. The potential is given by,
\vspace{3mm}
\bea\label{potentialsum}
V(b, \Psi) = (2\pi b)^3 \Lambda + \!\!\sum_{\rm particles}\!\!N_f\,E_{0}(b, \Psi, m)\ ,
\eea
where $\Lambda$ is the cosmological constant, $E_{0}$ is the Casimir energy for a scalar of mass $m$, and $N_f$ is the number of degrees of freedom, with a positive sign for bosons
and a negative sign for fermions (i.e., $N_f=2$ for the photon and graviton, $N_f = -4$ for a Dirac neutrino, $N_f = -2$ for a Majorana neutrino \cite{Arkani, Ponton, Rubin}). In formula (\ref{1Daction}) the matrix $\hat{M}$ has the following nonzero entries,
\vspace{3mm}
\bea
&&M_{11} = \frac{\rho_2^2+\rho_3^2}{2\tau_2^2\rho_3^2}\ , \ \ M_{22} =  \frac{3\rho_2^2+4\rho_3^2}{6\tau_2^2\rho_3^2}\ ,\ \ M_{55} = \frac{2}{3\rho_3^2}\ ,\nonumber\\
&&M_{25} = M_{52} = -\frac{1}{3\tau_2\rho_3}\ , \ \ M_{33} = M_{44} = \frac{1}{2\rho_3^2} \ , \nonumber\\
&&M_{13} = M_{31} = M_{24} = M_{42} = \frac{\rho_2}{2\tau_2\rho_3^2}\ .
\eea
It is easy to check that $\hat{M}$ is positive definite. Varying the action (\ref{1Daction}) with respect to $N$ and setting $N=1$ (which corresponds to fixing the gauge) we arrive at,
\bea\label{eqnow2}
\frac{1}{2}M_{p}^2(2\pi)^3\left(-6\,b\,\dot{b}^2 + b^3\,\dot{\Psi}^{\rm T} \hat{M} \dot{\Psi}
\right)+ V(b, \Psi)=0\ ,
\eea
thus the total energy has to vanish. As a consequence, the existence of a vacuum at $(b_0, \Psi_0)$ requires $V(b_0, \Psi_0)=0$.
In addition, we can set $N=1$ directly in the action (\ref{1Daction}) and write down the equations of motion that arise from varying the action
with respect to the other parameters. For the
volume modulus it takes the form,
\bea\label{eqnow}
\frac{\ddot{b}}{b} + \frac{1}{2}\frac{\dot{b}^2}{b^2} +\frac{1}{4}\dot{\Psi}^{\rm T} \hat{M} \dot{\Psi} -\frac{1}{48\pi^3 M_{p}^2 b^2}\frac{\partial V(b, \Psi)}{\partial b}=0\ .
\eea
As noted in \cite{Arkani}, since all shape moduli have positive definite kinetic energy, while for the volume modulus it is negative, the conditions for the existence of a stable vacuum are,
\begin{eqnarray}\label{conditions}
V=0\ , \ \ \partial_{b} V =\partial_{\alpha} V =0\ , \ \ \partial_b^2 V < 0\ , \ \ \partial_\alpha^2 V > 0
\end{eqnarray}
at the stationary point, where $\alpha = \tau_1, \tau_2, \rho_1, \rho_2, \rho_3$.
This presents a fine tuning problem since both the potential and its derivative have to vanish at the same point.
In addition, as we will shortly show, even conditions (\ref{conditions}) themselves cannot be fulfilled simultaneously.

Note that the potential $V(b, \Psi)$ is expressed in terms of bare quantities, each of which is divergent. We first write the cosmological constant as,
\begin{eqnarray}
\Lambda = \Lambda^{\rm obs} + \Lambda^{\rm div}\ ,
\end{eqnarray}
where $\Lambda^{\rm obs}\simeq3.1\times 10^{-47}~{\rm GeV}^4$ \cite{pdg} is the observed value,
and $\Lambda^{\rm div}$ is the divergent quantum correction, equal to the sum of
Casimir energies of particles
in flat space,
\begin{eqnarray}
  \Lambda^{\rm div} =\frac{\Gamma\left(-2\right)}{32\,\pi^2}
  \sum_{\rm particles}N_f \, m^4\ .
\end{eqnarray}
The Casimir energy for a scalar of mass $m$ in a 4D spacetime with spatial dimensions compactified on a 3-torus, assuming periodic boundary conditions, is,
\bea\label{sum}
E_0(b, \Psi, m) = \frac{1}{2}\sum_{n_1, n_2, n_3 = -\infty}^\infty \left(T^{i j}n_i n_j+m^2\right)^{\frac{1}{2}},
\eea
where $T^{ij}$ is the inverse of $T_{ij}$ given by equation (\ref{torus}).
The regularized expression for the triple sum in (\ref{sum}) is derived in the appendix. We immediately notice that the divergent parts in formula (\ref{potentialsum}) cancel and we can write the potential as,
\bea\label{pot}
V(b, \Psi) = (2\pi b)^3 \Lambda^{\rm obs} + \!\!\sum_{\rm particles}N_f\,E_{0}^{\rm obs}(b, \Psi, m)\ ,
\eea
where the finite part of the Casimir energy (\ref{sum}) is given by,
\bea\label{CE}
\lefteqn{\,E_0^{\rm obs}(b, \Psi, m) = -\frac{1}{\pi}\frac{1}{\sqrt{T^{11}}} \Bigg\{m \,\sqrt{T^{11}}\sum_{n=1}^\infty \frac{1}{n} \, K_1\!\left(\tfrac{2\pi m}{\sqrt{T^{11}}}n\right) }\nonumber\\
& & \!+\,  \sqrt{T^{11}}\!\!\!\!\!\!\sum_{n_2, n_3 = -\infty}^\infty\!\!\!\!\!\!\!'\ \ \ \,\sum_{n_1=1}^\infty \frac{1}{n_1}\cos\Big[\tfrac{2\pi}{T^{11}}n_1 (n_2 T^{12}+n_3 T^{13})\Big]\nonumber\\
& & \ \ \,\times \,\sqrt{d(n_2, n_3)+m^2}\,\,\,K_1\!\left[\tfrac{2\pi}{\sqrt{T^{11}}}n_1\sqrt{d(n_2, n_3)+m^2}\right] \nonumber\\
& & \!+\,\,m^{3/2}\,\Delta_{11}^{1/4}\sum_{n=1}^\infty \frac{1}{n^{3/2}}\, K_{3/2}\!\left(\tfrac{2\pi \, m}{\sqrt{\Delta_{11}}}\,n\right)\nonumber\\
& & \!+\,\, m^{2}\,\sqrt{\frac{D'}{\Delta_{11}}}\, \sum_{n=1}^\infty \frac{1}{n^2}\,K_{2}\!\left(\tfrac{2\pi\, m}{\sqrt{D'}}\,n\right)\nonumber\\
& & \!+\,2\,\Delta_{11}^{1/4}\!\!\sum_{n_2, n_3=1}^\infty \left(D' n_3^2 + m^2\right)^{3/4} \cos\left[2\,\pi\,n_2\,n_3\tfrac{\Delta_{12}}{\Delta_{11}}\right]\nonumber\\
& & \ \ \,\times \,\frac{1}{n^{3/2}_2}\,K_{3/2}\!\left(\tfrac{2\pi}{\sqrt{\Delta_{11}}}n_2\sqrt{D' n_3^2+m^2}\right)\Bigg\}\ .
\eea
In formula (\ref{CE}), $K_n(x)$ is the modified Bessel function of the second kind, the matrix $\hat{\Delta}$ and function $d$ are,
\vspace{3mm}
\bea
\!\!\hat{\Delta} = \frac{1}{T^{11}}\Bigg(\!
  \begin{array}{cc}
    T^{11} T^{22}-(T^{12})^2 \ &\  T^{11} T^{23}-T^{12}T^{13} \\
    T^{11} T^{23}-T^{12}T^{13} \ &\  T^{11} T^{33}-(T^{13})^2 \\
  \end{array}\!
\Bigg),
\eea
\vspace{-4mm}
\bea
d(n_2, n_3) = \left(\!
                                                                                         \begin{array}{cc}
                                                                                           n_2 & n_3 \\
                                                                                         \end{array}\!
                                                                                       \right)
\hat{\Delta}\left(\!
                                                                                    \begin{array}{c}
                                                                                      n_2 \\
                                                                                      n_3 \\
                                                                                    \end{array}\!
                                                                                  \right),
\eea
and $D' = \det(\hat{\Delta})/\Delta_{11}$.
In the massless limit formula (\ref{CE}) reduces to,
\vspace{1mm}
\bea\label{CE0}
\lefteqn{\ E_0^{\rm obs}(b, \Psi, 0) \ =\  -\frac{1}{\pi}\frac{1}{\sqrt{T^{11}}}\ \Bigg\{\frac{\pi}{12} \ T^{11} }\nonumber\\
& & \!+\,  \sqrt{T^{11}}\!\!\!\!\!\!\sum_{n_2, n_3 = -\infty}^\infty\!\!\!\!\!\!\!'\ \ \ \,\sum_{n_1=1}^\infty \frac{1}{n_1}\cos\Big[\tfrac{2\pi}{T^{11}}n_1 (n_2 T^{12}+n_3 T^{13})\Big]\nonumber\\
& & \ \ \,\times \,\sqrt{d(n_2, n_3)}\,\,\,K_1\!\left[\tfrac{2\pi}{\sqrt{T^{11}}}n_1\sqrt{d(n_2, n_3)}\right] \nonumber\\
& & \!+\,\frac{\zeta(3)}{4\pi}\Delta_{11} + \frac{\pi^2}{180}\frac{1}{\sqrt{\Delta_{11}}}D'^{\ 3/2}\nonumber\\
& & \!+\,2\,\Delta_{11}^{1/4}\,D'^{\ 3/4}\sum_{n_2, n_3=1}^\infty  \cos\left[2\,\pi\,n_2\,n_3\tfrac{\Delta_{12}}{\Delta_{11}}\right]\nonumber\\
& & \,\ \  \times \,
\left(\frac{n_3}{n_2}\right)^{3/2}K_{3/2}\!\left(2\,\pi \,n_2 \,n_3\,\sqrt{\tfrac{D'}{\Delta_{11}}}\right)\Bigg\}\ .
\eea
\vspace{1mm}

Note that for $m \gg 1/b$ the Casimir energy (\ref{CE}) behaves like  $\exp(- C\, b \,m)$, where $C$ is a constant and depends on the  shape moduli.
We restrict our attention to the lengthscale $b \gg 1/m_e $, so that the Casimir energies of the electron and all heavier standard model particles are negligible compared to the contributions of the photon, graviton, and neutrinos.\footnote{Our results hold for the full range of $b$ where the standard model is valid (see figures 1 and 2).}

It turns out that even before performing the numerical analysis, we can precisely determine the values of the shape moduli  for which the potential (\ref{pot}) has its extrema. It can be shown that the Casimir energy (\ref{sum}) is invariant under $\textrm{SL}(3, \mathbb{Z})$ transformations. The nine generators of the $\textrm{SL}(3, \mathbb{Z})$ group are listed in \cite{SL3,generators}.
For example, the generator  $T_1:\tau_1\rightarrow\tau_1+1$ corresponds to a change of indices $(n_1, n_2, n_3) \rightarrow (n_1, n_2-n_1, n_3)$ in (\ref{sum}), whereas
$T_3:\rho_3\rightarrow \rho_3+1$ is equivalent to replacing $(n_1, n_2, n_3) \rightarrow (n_1, n_2, n_3-n_1)$. The same symmetries are exhibited by the potential (\ref{pot}), since it is a linear combination of Casimir energies of the particles.
It has been argued that fixed points of the transformation under which the potential is invariant
correspond to extrema of this potential \cite{6D4,Ponton,Wilczek}. Such fixed points should also lie on the boundary of the fundamental
domain of the symmetry group. The fundamental region for a 3-torus parametrized as in (\ref{torus}) is the following \cite{SL3,fund},
\vspace{2mm}
\bea\label{fundamental}
&&\!\!\!\!\!\!\!\!1 \leq \tau_1^2+\tau_2^2 \leq \rho_1^2+\rho_2^2+\rho_3^2\ , \ \ \  -1/2 < \rho_1, \tau_1 \leq 1/2\ ,  \nonumber\\
&&\ \ \rho_1\tau_1 + \rho_2\tau_2 \le (\tau_1^2+\tau_2^2)/2\ , \ \ \ \ \ \ \tau_2 > 0\ .
\eea
This is the moduli space of physically distinct 3-tori.
Fixed points of $\textrm{SL}(3, \mathbb{Z})$ correspond to the case when the inequalities in the first and third relation in (\ref{fundamental}) become equalities, while $\tau_1, \rho_1$ are $0$ or $1/2$.
A numerical analysis shows that the fixed point corresponding to a minimum of the potential exists for $\tau_1 = \rho_1 = 1/2$, thus the shape moduli for a vacuum stable in the subspace $\left(\tau_1, \tau_2, \rho_1, \rho_2, \rho_3\right)$ are,
\vspace{2mm}
\bea\label{shape}
\Psi_0^{\rm T} = \left(\frac{1}{2}, \frac{\sqrt{3}}{2}, \frac{1}{2}, \frac{\sqrt{3}}{6}, \frac{\sqrt{6}}{3}\right)\ .
\eea

Although neutrino masses have not been determined, we can use experimental mass splittings for the atmospheric and solar neutrinos to generate the spectrum given the lightest neutrino mass and a choice of hierarchy. This allows us to investigate the potential for various lightest neutrino masses. Experimentally, $\Delta m^2_{\rm atm} = (2.43 \pm 0.13)\times 10^{-3} {\ \rm eV^2}$, $\Delta m^2_{\rm sol} = (7.59\pm 0.20)\times 10^{-5} {\ \rm eV^2}$ \cite{pdg}.
Denoting the lightest neutrino mass by $m_l$, the masses of the other two neutrinos, assuming normal hierarchy, are
$m_{l}^2+\Delta m^2_{\rm sol}$ and
$m_{l}^2+\Delta m^2_{\rm atm}+\Delta m^2_{\rm sol}$, whereas
for an inverted hierarchy the masses are
$m_{l}^2 + \Delta m^2_{\rm atm} - \Delta m^2_{\rm sol}$ and
$m_{l}^2+\Delta m^2_{\rm atm}$.

\begin{figure}[t]
\vspace{2mm}
\centerline{\scalebox{1.07}{\includegraphics{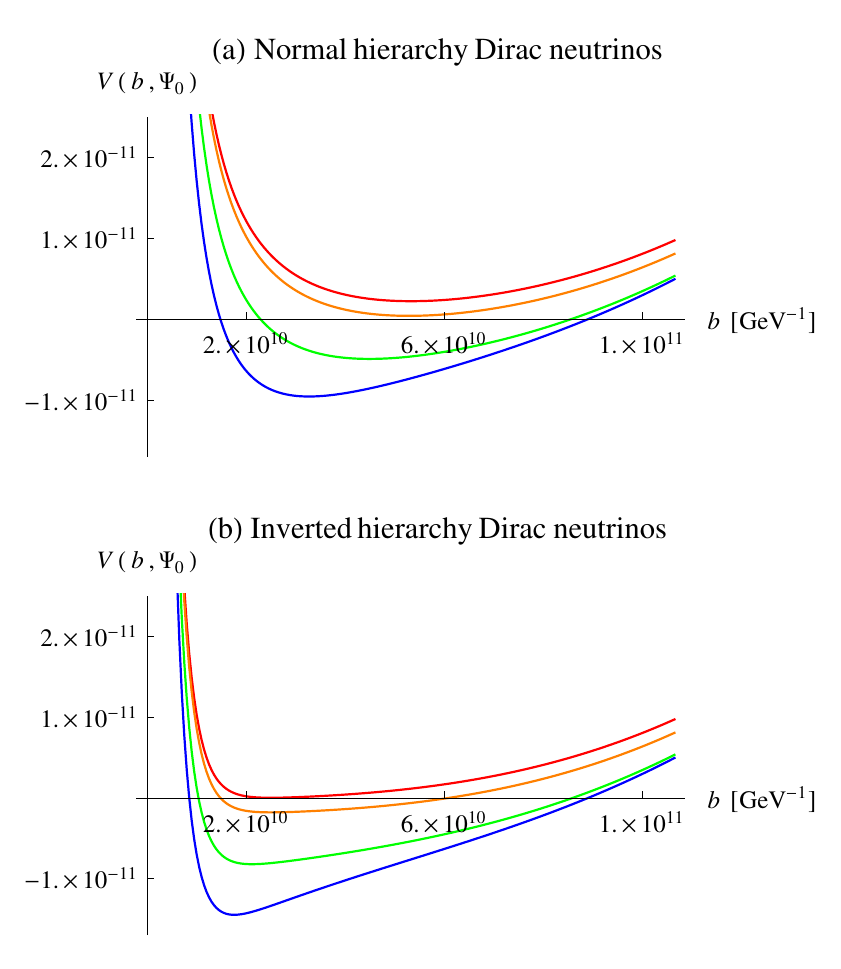}}}
\vspace{-2mm}
\caption{\footnotesize{(a) Plots of $V(b, \Psi_0)$ for Dirac neutrinos with normal hierarchy for masses $m_{l} = 0$ (red),
$10^{-12}\ {\rm GeV}$ (orange),  $5\times 10^{-12}\ {\rm GeV}$ (green), and  $10^{-11}\ {\rm GeV}$ (blue). (b) The same for an inverted hierarchy. }}
\end{figure}
Under our assumption $b \gg 1/m_e$, the potential in case of the standard model with Dirac neutrinos is,
\bea\label{10}
\lefteqn{\!\!\!\!V(b, \Psi_0) = (2\pi b)^3  \Lambda^{\rm obs}}\nonumber\\
& & + \bigg[\,4\,E^{\rm obs}(b, \Psi_0, 0)   - \,4\,\sum_{i=1}^3 E^{\rm obs}(b, \Psi_0, m_{\nu_i}) \bigg].
\eea

\vspace{5mm}
\hspace{-2mm}The plots of $V(b, \Psi_0)$ for  several lightest neutrino masses for normal and inverted hierarchy Dirac neutrinos are given in figure 1 (a) and (b), respectively. The only extremum of the potential is a minimum, but the conditions for a stable stationary point (\ref{conditions}) require it to be a maximum. This proves that there are no stable one-dimensional vacua of the low-energy effective theory.
Nevertheless, we find precisely one set of neutrino masses for each type of hierarchy for which an unstable vacuum exists. In the case of normal hierarchy Dirac neutrinos the lightest neutrino mass for such an unstable vacuum is  $m_l \approx 10^{-12} {\rm \ GeV}$, whereas in the inverted hierarchy case it is $m_l \approx 0$. Both unstable vacua appear at the micron scale.

In the case of the standard model with Majorana neutrinos the potential takes the form,
\bea\label{11}
\lefteqn{\!\!\!\!V(b, \Psi_0) = (2\pi b)^3  \Lambda^{\rm obs}}\nonumber\\
& & + \bigg[\,4\,E^{\rm obs}(b, \Psi_0, 0)   - \,2\,\sum_{i=1}^3 E^{\rm obs}(b, \Psi_0, m_{\nu_i}) \bigg].
\eea
Figure 2 (a) and (b) shows the plot of $V(b, \Psi_0)$ for Majorana neutrinos for a few lightest neutrino masses. Note that in this case there does not even exist an unstable
vacuum.
\begin{figure}[t]
\vspace{-4mm}
\centerline{\scalebox{1.07}{\includegraphics{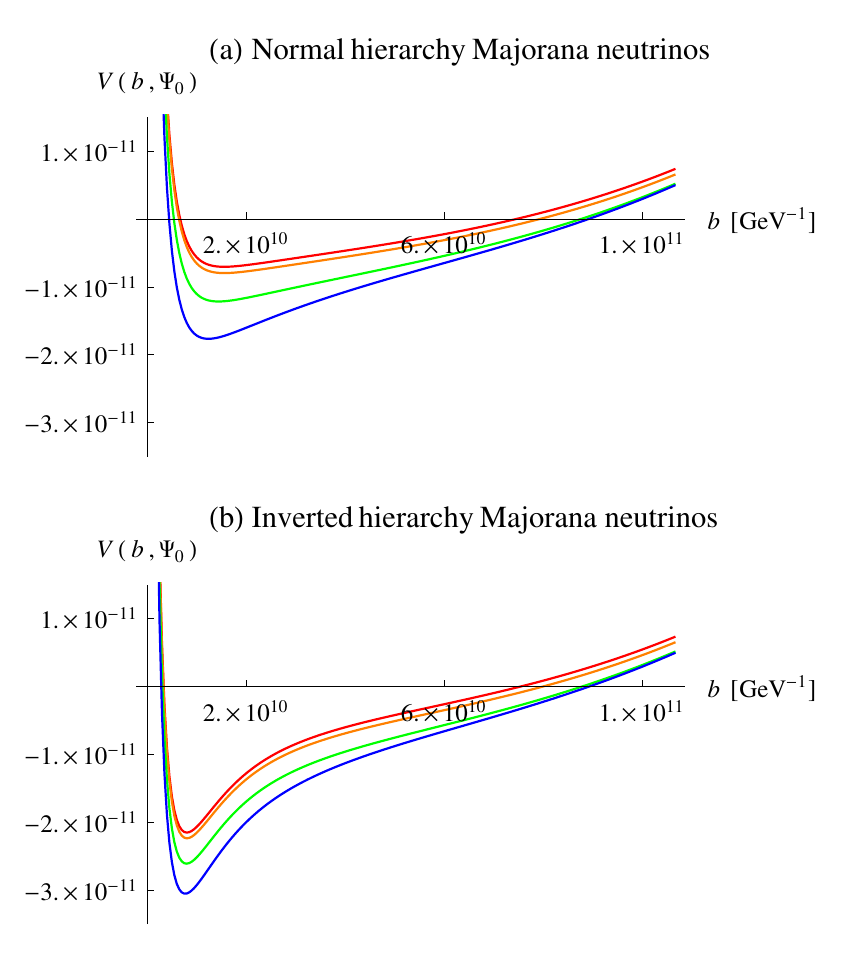}}}
\vspace{-2mm}
\caption{\footnotesize{Plots of $V(b, \Psi_0)$ for Majorana neutrinos with normal hierarchy (figure (a)) and inverted hierarchy (figure (b)) for masses $m_{l} = 0$ (red),
$10^{-12}\ {\rm GeV}$ (orange),  $5 \times 10^{-12}\ {\rm GeV}$ (green), and  $10^{-11}\ {\rm GeV}$ (blue).}}
\end{figure}

\vspace{2cm}

\section{Compactifications on other 3D manifolds}
Our analysis from the last section can be easily extended to  other topologies of the compact space, for instance
$S^1\times S^1\times S^1$, $S^1\times T^2$, $S^3$, and $S^1\times S^2$.
The first two cases are very similar to $T^3$. We briefly comment on the
other two possibilities, which are considerably different because of a nonzero curvature
of the compact space.

\subsection{Compactification on $S^1\times S^1\times S^1$}
Denoting the radii of compactification by $R_1, R_2, R_3$, the metric takes the form,
\bea\label{three_circles}
\!\!\!ds^2 = - N^2 d t^2 + R_1^2 (d y^1)^2 + R_2^2 (d y^2)^2 + R_3^2 (d y^3)^2,
\eea
where \,$y^1, y^2, y^3 \in [ 0,  2\pi )$. The dimensionally reduced action (\ref{4Daction}) is,
\begin{eqnarray}\label{1Daction2}
S = \!\int \!d t \left[\frac{1}{2}M_{p}^2\frac{\rm Vol_1}{N}\dot{\Phi}_1^{\rm T}\hat{S}\,\dot{\Phi}_1-N\, V_1(\Phi_1)\right],
\end{eqnarray}
with $\,\Phi_1^{\rm T} = (\log R_1, \log R_2, \log R_3)$, ${\rm Vol}_1 = (2\pi)^3 R_1 R_2 R_3$, and the potential,
\bea\label{potentialsumCIRCLE}
\!\!V_1(\Phi_1)= {\rm Vol}_1 \ \Lambda^{\rm obs} + \!\!\!\sum_{\rm particles}\!\!N_f\,E_{1}^{\rm obs}(R_1, R_2, R_3, m).
\eea
The only nonzero elements of matrix $\hat{S}$ are,
\bea
S_{12} = S_{21} = S_{23} = S_{32} = S_{13} = S_{31} = -1\ .
\eea
The Casimir energy for a scalar particle of mass $m$ is calculated using formula (\ref{for20}) from the appendix with the appropriate choice of metric
and is given by,
\bea\label{CE2}
\lefteqn{\!\!\!\!\!\!\!\!\!\!E_{1}^{\rm obs}(R_1, R_2, R_3, m) = -\frac{1}{\pi}R_1\,\Bigg\{\frac{m}{R_1}\,\sum_{n=1}^\infty \frac{1}{n} \,K_{1}\!\left(2\pi\, m\,R_1\,n\right) }\nonumber\\
& & \!+\,\, \frac{1}{R_1}\sum_{n_2, n_3 = -\infty}^\infty\!\!\!\!\!\!\!'\ \ \ \,\sum_{n_1=1}^\infty \frac{1}{n_1}\sqrt{(\tfrac{n_2}{R_2})^2 + (\tfrac{n_3}{R_3})^2 +m^2}\nonumber\\
& & \ \ \,\,\times \,K_{1}\left[2\pi R_1n_1\sqrt{(\tfrac{n_2}{R_2})^2 + (\tfrac{n_3}{R_3})^2 +m^2}\right]\nonumber\\
& & \!+\,\,m^{3/2}\frac{1}{\sqrt{R_2}}\sum_{n=1}^\infty \frac{1}{n^{3/2}} \,K_{3/2}\!\left(2\pi\, m\, R_2\,n\right)\nonumber\\
& & \!+\,\, m^{2}\frac{R_2}{R_3}\sum_{n=1}^\infty \frac{1}{n^2}\,K_{2}\!\left(2\pi \,m\, R_3\,n\right)\nonumber\\
& & \!+\,\,\frac{2}{\sqrt{R_2}}\sum_{n_2, n_3=1}^\infty \frac{1}{n_2^{3/2}}\left[(\tfrac{n_3}{R_3})^2 + m^2\right]^{3/4} \nonumber\\
& & \ \ \,\,\times \,K_{3/2}\left(2\pi\,R_2\,n_2\sqrt{(\tfrac{n_3}{R_3})^2+m^2}\right)\Bigg\}\ .
\eea
Note that the potential is invariant under the permutation of $(R_1, R_2, R_3)$, which is not obvious from formula (\ref{CE2}).
Numerical analysis reveals that the only extremum of the potential is a minimum.
The same reasoning as in the 3-torus case leads to a vanishing potential  at the stationary point, which is accomplished again only for Dirac neutrinos,
at $R_1 = R_2 = R_3 \approx 3\times 10^{10} {\rm \ GeV^{-1}}$, $m_l \approx 10^{-12} {\rm \ GeV}$ in case of normal hierarchy, and at
$R_1 = R_2 = R_3 \approx 5\times 10^{10} {\rm \ GeV^{-1}}$, $m_l \approx 0$ for inverted hierarchy. The conditions fulfilled at the only possible candidate for a stationary point are, therefore,
\begin{eqnarray}\label{conditionsCIRCLE}
V=0\ , \ \ \partial_{\alpha} V  = 0\ , \ \ \partial_\alpha^2 V < 0\ ,
\end{eqnarray}
where $\alpha = R_1, R_2, R_3$.
Unfortunately, the matrix $\hat{S}$ is not positive definite, which indicates that
the existing stationary point is not stable.
Thus, the compactification on the manifold $S^1\times S^1\times S^1$ does not differ qualitatively from the 3-torus case and there is only
one unstable vacuum for Dirac neutrinos for each choice of hierarchy.

\subsection{Compactification on $S^1\times T^2$}
In this case the metric is given by,
\bea\label{three_circles}
ds^2 = - N^2 d t^2 + R^2 (d y^1)^2 + t_{i j} d y^i d y^j\ ,
\eea
where,
\bea
t_{i j} = \frac{b^2}{\tau_2}\left(
            \begin{array}{cc}
              1 & \tau_1 \\
              \tau_1 & \tau_1^2+\tau_2^2 \\
            \end{array}
          \right),
\eea
$i, j = 2, 3$ and $y^1, y^2, y^3 \in [0, 2\pi)$.
The  reduced action is,
\begin{eqnarray}\label{1Daction2}
S = \int \!d t \,\bigg[\frac{1}{2}M_{p}^2\frac{\rm Vol_2}{N}\left(\dot{\Phi}_2^{\rm T} \hat{K} \,\dot{\Phi}_2
\right)-N \,V_2(\Phi_2)\bigg],
\end{eqnarray}
where $\,\Phi_2^{\rm T} = (\log R, \log b, \tau_1, \tau_2)$, ${\rm Vol}_2 = (2\pi)^3 R\, b^2$,  and the nonzero entries of $\hat{K}$ are,
\bea
\!\!\!K_{12} = K_{21} = -2\ , \ \ K_{22} = -2\ , \ \ K_{33} = K_{44} = \frac{1}{2\tau_2^2}\ .
\eea
As was discussed in \cite{AFW}, two-dimensional vacua for the compactification on a 2-torus are characterized by the shape moduli $(\tau_1, \tau_2) = (1/2, \sqrt{3}/2)$. In the $S^1\times T^2$ case we find that those  values also correspond to a minimum of the potential.
Since in the $(\tau_1, \tau_2)$ subspace the matrix $\hat{K}$ is positive definite, the above parameters describe a point stable in the directions $(\tau_1, \tau_2)$.
Nevertheless, the subspace $(R, b)$ of matrix $\hat{K}$ is not positive definite. Since the only existing stationary point of $V_2(R, b, 1/2, \sqrt{3}/2)$ is a minimum in both $R$ and $b$, it necessarily corresponds to an unstable vacuum, and appears again only for Dirac neutrinos at $R \approx b \approx 3\times 10^{10} {\rm \ GeV^{-1}}$, $m_l \approx 10^{-12} {\rm \ GeV}$ for normal hierarchy, and $R \approx b \approx 6\times 10^{10} {\rm \ GeV^{-1}}$, $m_l \approx 0$ for inverted hierarchy.

\subsection{Compactification on $S^3$}
For the compactification on a sphere the metric is,
\bea\label{three_circles}
\!\!\!ds^2 \!=\! - N^2 d t^2 + R^2\left[d \theta^2\! +\! \sin^2\theta \left(d \psi^2 \!+ \sin^2\!\psi \,d \phi^2\right)\right],
\eea
where \,$\theta, \psi \in [0, \pi)$ and $\phi \in [0, 2\pi)$.
The reduced action is,
\begin{eqnarray}\label{1Daction2}
S =  \int \!d t \left[-M_{p}^2\frac{6\pi^2}{N}R{\dot{R}^2} -N \,V(R)\right],
\end{eqnarray}
with the potential given in terms of finite quantities,
\bea
\!\!\!V(R)\! = \!2\pi^2R^3\!\left(\!-\frac{3 M_{p}^2}{R^2}\!+\!\Lambda^{\rm obs}\!\right)\! +\! \!\!\!\sum_{\rm particles}\!\!\!N_f E_{3}^{\rm obs}(R, m)\,.
\eea
Similar arguments as before yield the conditions at the stationary point,
\bea\label{spherecond}
V = 0 \ , \ \ \partial_R V = 0\ .
\eea
Note that this case is qualitatively different from the previous ones because of a nonzero curvature term.
We find that Casimir energies are negligible compared to this curvature term for $R \gg 1/M_{p}$, which is well satisfied in the region
we are considering ($R \gg 1/m_e$). It is now straightforward to check that both conditions (\ref{spherecond}) cannot be fulfilled
simultaneously, which proves that there are no one-dimensional vacua. This remains true even after introducing a magnetic flux (see \cite{AFW} for how this argument works in case of a two-dimensional compactification on a sphere).
Choosing the compact topology to be $S^1\times S^2$ yields exactly the same conclusions.

We have also analyzed 3D compactifications on surfaces of genus greater than one.
For analogous reasons as those presented in \cite{AFW}, no vacua exist in those cases.

\section{Conclusions}
We have investigated the  structure of the standard model coupled to
gravity with spatial dimensions
compactified on three-dimensional manifolds.
We have focused on the 3-torus compactification, as it seems the most
natural three-dimensional
topology with no curvature. Other cases can be explored in a
similar fashion.

For the 3-torus case, we have
analyzed the standard model with Dirac and Majorana neutrinos,
both for normal and inverted hierarchy.
We have calculated the effective potential, which contains, apart from the cosmological constant term,
the  Casimir energies of the graviton, photon and neutrinos.  The Casimir energies of particles of higher mass  are negligible.
We have found, arguing on the basis of the symmetry exhibited by the potential, the unique choice of the toroidal shape parameters required to
have a stable vacuum in this subspace. The potential then becomes a function of just the volume modulus and is
precisely determined by the shape moduli,
neutrino masses, and their type.
We have  shown that there are no stable vacua of the low-energy effective theory, since
the volume modulus has a negative kinetic term, while the only extremum of the effective potential
is a minimum.
Nevertheless, we have  found that
in case of Dirac neutrinos there exists an unstable one-dimensional vacuum
 for precisely one set of neutrino masses for
each type of hierarchy. The volume modulus for this unstable vacuum is on
the order of microns.  This stationary point disappears at high enough temperatures.

For the compactifications on $S^1\times S^1\times S^1$
and $S^1\times T^2$ similar conclusions were found. The cases with spatial
dimensions compactified
on $S^3$ or $S^1\times S^2$ differ qualitatively because of the
presence of a nonzero curvature term. We have shown that
there are no one-dimensional vacua in those cases. A similar conclusion
is reached for any compactification on a surface of genus greater than one.

\vspace{2mm}

\subsection*{Acknowledgment}
The work of the authors was supported in part by the U.S. Department of Energy under contract No. DE-FG02-92ER40701.

\vspace{5mm}

\appendix\label{app}
\section{\textit{Generalized multidimensional Chowla-Selberg formula}}
In this section we present a derivation of the formula for the regularized triple sum in equation (\ref{sum}). Some steps of this calculation are given in \cite{eli,elizalde3}. It can be shown \cite{AFW} that,
\bea\label{aa3}
\!\!\!\sum_{n=-\infty}^{\infty}\!\!\!e^{-(n+z)^2 w}= \sqrt{\frac{\pi}{w}}\left[1+2\sum_{n=1}^{\infty}e^{- \frac{\pi^2 n^2}{w}}\cos(2\pi n z)\right]
\eea
under the condition ${\rm Re}(w)>0$. We can also write,
\bea
\!\!\!\!\!\left(\vec{n}^{\rm T} \hat{A}\, \vec{n}  +q\right)^{-s} = \frac{1}{\Gamma(s)}\int_0^\infty dt \,t^{s-1}e^{-(\vec{n}^{\rm T} \hat{A}\, \vec{n} + q) t}\ ,
\eea
where $\vec{n}^{\rm T} = (n_1,\, n_2,\, n_3)$. We assume $A_{11}, q > 0$ and write the quadratic form as,
\bea
\!\!\!\!\vec{n}^{\rm T} \hat{A}\, \vec{n} = A_{11}\left(n_1+\tfrac{A_{12}}{A_{11}}n_2+\tfrac{A_{13}}{A_{11}}n_3\right)^2 + d(n_2, n_3)\ ,
\eea
with
\bea
\lefteqn{\!\!\!\!\!\!\!\!\!\!\!d(n_2, n_3) = \left(\!
                                                                                         \begin{array}{cc}
                                                                                           n_2 & n_3 \\
                                                                                         \end{array}\!
                                                                                       \right)
\hat{\Delta}\left(\!
                                                                                    \begin{array}{c}
                                                                                      n_2 \\
                                                                                      n_3 \\
                                                                                    \end{array}\!
                                                                                  \right) }\nonumber\\
& &\!\!\!\!\!\!\!\!\!\!\!\!\!\!\!\!\!\!\!=\left(\!
        \begin{array}{cc}
          n_2 & n_3 \\
        \end{array}\!
      \right)\left(\!
  \begin{array}{cc}
    A_{22}-\frac{A_{12}^2}{A_{11}} & A_{23}-\frac{A_{12}A_{13}}{A_{11}} \\
    A_{23}-\frac{A_{12}A_{13}}{A_{11}} & A_{33}-\frac{A_{13}^2}{A_{11}} \\
  \end{array}\!
\right)\left(\!
                                                                                    \begin{array}{c}
                                                                                      n_2 \\
                                                                                      n_3 \\
                                                                                    \end{array}\!
                                                                                  \right).
\eea
Using  relation (\ref{aa3}) with respect to the index $n_1$ we get,
\vspace{1mm}
\bea\label{aa6}
\lefteqn{\!\!\!\!\!\!\!\!\!\!\!\sum_{n_1, n_2, n_3=-\infty}^\infty\left(\vec{n}^{\rm T} \hat{A}\, \vec{n} +q\right)^{-s} }\nonumber\\
& & \!\!\!\!\!\!\!\!\!\!\!\!\!\!\!\!\!\! =\frac{1}{\Gamma(s)}\sqrt{\frac{\pi}{A_{11}}}\sum_{n_2, n_3=-\infty}^{\infty}\int_0^\infty dt\,t^{s-\frac{3}{2}}e^{-\left[d(n_1, n_2)+q\right]t}\nonumber\\
& & \!\!\!\!\!\!\!\!\!\!\!\!\!\times  \left[1+ 2\sum_{n_1=1}^\infty e^{-\frac{\pi^2 n_1^2}{ A_{11} t }}\cos\left[2\pi \, n_1\left(\tfrac{A_{12}n_2+A_{13}n_3}{A_{11}}\right)\right]\right]\!.
\eea
The $(n_2, n_3) = (0, 0)$ contribution to (\ref{aa6}) is,
\vspace{1mm}
\bea
q^{-s}+2\,A_{11}^{-s}\sum_{n_1=1}^\infty \left(n_1^2+\tfrac{q}{A_{11}}\right)^{-s}\ .
\eea
Now, making use of the following property of modified Bessel functions of the second kind,
\vspace{1mm}
\bea
\int_0^\infty du\,u^{s-1}e^{-\alpha^2 u - \frac{\beta^2}{ u}} = 2\left(\tfrac{\beta}{\alpha}\right)^s K_s(2\,\alpha\,\beta)\ ,
\eea
the $(n_2, n_3) \ne (0, 0)$ contribution to (\ref{aa6}) is,
\vspace{1mm}
\bea\label{aaa8}
\lefteqn{\!\!\!\!\!\frac{\Gamma(s-\tfrac{1}{2})}{\Gamma(s)}\sqrt{\frac{\pi}{A_{11}}} \sum_{n_2, n_3=-\infty}^\infty \!\!\!\!\!\!\!'\ \ \left[d(n_2, n_3) + q\right]^{-s+\frac{1}{2}} }\nonumber\\
& &\!\!\!\!\!\!\!\!\!\!\!\!+\,\frac{4\,\pi^s}{\Gamma(s)}A_{11}^{-\frac{s}{2}-\frac{1}{4}}\!\!\sum_{n_2, n_3=-\infty}^\infty\!\!\!\!\!\!\!'\ \ \ \ \sum_{n_1=1}^\infty
\left[ d(n_2, n_3) + q \right]^{-\frac{s}{2}+\frac{1}{4}}\nonumber\\
& & \!\!\!\!\!\!\! \times\  n_1^{s-\frac{1}{2}}\cos\left[\tfrac{2\pi n_1}{A_{11}}\,(A_{12}n_2+A_{13}n_3)\right]\nonumber\\
& & \!\!\!\!\!\!\! \times\  K_{s-\frac{1}{2}}\left(\tfrac{2\pi n_1}{\sqrt{A_{11}}}\sqrt{d(n_2, n_3)+q}\right),
\eea
where the prime indicates excluding the $(0, 0)$ term.
In order to calculate the first term in (\ref{aaa8}) we use the result of \cite{AFW} and, under the assumptions $\Delta_{11},  \det(\hat{\Delta}) > 0$, write the sum over $n_2$ and $n_3$ as,
\bea\label{app20}
\lefteqn{\!\!\!\!\!\!\!\!\!\!\sum_{n_2, n_3=-\infty}^\infty\!\!\!\!\!\!\!'\ \left[d(n_2, n_3) +q\right]^{-s+\frac{1}{2}} = 2\,\Delta_{11}^{-s+\frac{1}{2}}\zeta_{\rm EH}\left(s-\tfrac{1}{2}, \tfrac{q}{\Delta_{11}}\right) }\nonumber\\
&& \!\!\!\!\!\!\!\!\!\!\!\!\!\!\!+\, 2\,\sqrt{\pi}\,\frac{\Gamma(s-1)}{\Gamma\!\left(s-\frac{1}{2}\right)}\,\frac{\Delta_{11}^{s-\frac{3}{2}}}{D^{s-1}}\,\zeta_{\rm EH}\left(s-1, \tfrac{\Delta_{11}\,q}{D}\right) \nonumber\\
&& \!\!\!\!\!\!\!\!\!\!\!\!\!\!\!+\,\frac{8\,\pi^{s-\frac{1}{2}}}{\Gamma\!\left(s-\frac{1}{2}\right)}\frac{1}{\sqrt{\Delta_{11}}}\sum_{n_2, n_3=1}^{\infty} n_2^{s-1}
\left( D\,n_3^2 +\Delta_{11} \,q\right)^{-\frac{s}{2}+\frac{1}{2}}\nonumber\\
&& \!\!\!\!\!\!\!\!\!\!\!\!\!\! \times\cos\left(2\pi\,n_2\,n_3\,\tfrac{\Delta_{12}}{\Delta_{11}}\right)K_{s-1}\!\left(\tfrac{2\,\pi\,n_2}{\Delta_{11}}\sqrt{D\,n_3^2 + \Delta_{11}\,q}\right)
\eea
where $D = \det(\hat{\Delta})$ and the regularized form of the Epstein-Hurwitz zeta function is,
\bea\label{A4}
\lefteqn{\!\!\!\!\!\!\!\!\!\!\!\!\!\!\!\!\!\!\!\!\!\zeta_{\rm EH}\left(s, q\right) \!\equiv\! \sum_{n=1}^\infty\left(n^2+q\right)^{-s}  \!=\! -\tfrac{1}{2}q^{-s} \!+\! \frac{\sqrt{\pi}}{2}\frac{\Gamma(s-\frac{1}{2})}{\Gamma(s)}q^{-s+\frac{1}{2}} }\nonumber\\
& & +\,
\frac{2\,\pi^s}{\Gamma(s)}q^{\frac{1-2s}{4}}\sum_{n=1}^\infty n^{s-\frac{1}{2}}K_{s-\frac{1}{2}}(2\pi n \sqrt{q})\ .
\eea
The final formula for the regularized triple sum is, therefore,
\bea\label{for20}
\lefteqn{\!\!\sum_{n_1, n_2, n_3=-\infty}^\infty\!\!\!\!(\vec{n}^{\rm T} \hat{A}\, \vec{n} +q)^{-s} =\frac{\pi^{\frac{3}{2}}\,\Gamma(s-\frac{3}{2})}{\Gamma(s)\,\sqrt{A_{11}}\,\sqrt{D}}\,q^{-s+\frac{3}{2}} }\nonumber\\
& & \!\!\!\!\!\!\!+\,\frac{4\pi^s}{\Gamma(s)}\frac{1}{\sqrt{A_{11}}}\Bigg\{q^{-\frac{s}{2}+\frac{1}{4}}A_{11}^{-\frac{s}{2}+\frac{1}{4}}\sum_{n=1}^\infty n^{s-\frac{1}{2}}K_{s-\frac{1}{2}}\left(\tfrac{2\pi \sqrt{q}}{\sqrt{A_{11}}}n\right)\nonumber\\
& & \!+\, A_{11}^{-\frac{s}{2}+\frac{1}{4}}\!\!\sum_{n_2, n_3 = -\infty}^\infty\!\!\!\!\!\!\!'\ \ \ \,\sum_{n_1=1}^\infty n_1^{s-\frac{1}{2}}\left[d(n_2, n_3)+q\right]^{-\frac{s}{2}+\frac{1}{4}}\nonumber\\
& & \ \ \,\times \,K_{s-\frac{1}{2}}\left[\tfrac{2\pi}{\sqrt{A_{11}}}n_1\sqrt{d(n_2, n_3)+q}\right]\nonumber\\
& & \ \ \,\times \,\cos\left[\tfrac{2\pi}{A_{11}}n_1(n_2A_{12}+n_3A_{13})\right] \nonumber\\
& & \!+\,q^{-\frac{s}{2}+\frac{1}{2}}\Delta_{11}^{-\frac{s}{2}}\sum_{n=1}^\infty n^{s-1}K_{s-1}\left(\tfrac{2\pi \sqrt{q}}{\sqrt{\Delta_{11}}}\,n\right)\nonumber\\
& & \!+\, q^{-\frac{s}{2}+\frac{3}{4}}\frac{1}{\sqrt{\Delta_{11}}}D'^{-\frac{s}{2}+\frac{1}{4}}\sum_{n=1}^\infty n^{s-\frac{3}{2}}\,K_{s-\frac{3}{2}}\left(\tfrac{2\pi \sqrt{q}}{\sqrt{D'}}n\right)\nonumber\\
& & \!+\,2\,\Delta_{11}^{-\frac{s}{2}}\sum_{n_2, n_3=1}^\infty \left(D' n_3^2 + q\right)^{-\frac{s}{2}+\frac{1}{2}} \cos\left[2\,\pi\,n_2\,n_3\tfrac{\Delta_{12}}{\Delta_{11}}\right]\nonumber\\
& & \ \ \,\times \,n_2^{s-1}\,K_{s-1}\left(\tfrac{2\pi}{\sqrt{\Delta_{11}}}n_2\sqrt{D'  n_3^2+q}\right)\Bigg\}\ ,
\eea
where $D' = \det(\hat{\Delta}) / \Delta_{11}$.
In order to obtain the regularized formula for the Casimir energy density (\ref{sum}) we simply set,
\bea
\hat{A} = \hat{T}^{-1}\ , \ \ q=m^2\ , \ \ s=-\tfrac{1}{2}\ .
\eea
It can be checked that all our assumptions are then fulfilled. Thus, formula (\ref{for20}) applies and we arrive at equation (\ref{CE}).


\end{document}